\documentclass[
    reprint,
    superscriptaddress,
    amsmath,
    amssymb,
    aps,
    prd,
    nofootinbib,
    floatfix
]{revtex4-2}

\usepackage{graphicx}   
\usepackage{dcolumn}    
\usepackage{bm}         
\usepackage{xcolor}     
\usepackage{hyperref}   
\usepackage{cleveref}   
\usepackage{enumerate}
\usepackage{relsize}
\usepackage{tabularx}
\usepackage{empheq}
\usepackage[mathlines]{lineno}
\usepackage{orcidlink}
\usepackage[normalem]{ulem}

\allowdisplaybreaks

\hypersetup{
    colorlinks=true,
    citecolor=blue,
    filecolor=blue,
    linkcolor=blue,
    urlcolor=blue
}

\newcommand{\hmu}{\hat{\mu}}

\begin{document}


\title{A 4D $T^\prime$-expanded lattice QCD equation of state with a multi-dimensional \\ 3D-Ising-type critical contribution}

\author{Francesco~Di~Clemente\,\orcidlink{0000-0002-8257-3819}}
\email{fdicleme@central.uh.edu}
\affiliation{
 Department of Physics, University of Houston, Houston, TX 77204, USA
}

\author{Prachi Garella\,\orcidlink{0000-0002-9980-4855}}
\affiliation{
 Department of Physics, University of Houston, Houston, TX 77204, USA
}

\author{Johannes Jahan\,\orcidlink{0009-0002-4557-4652}}
\affiliation{
 Department of Physics, University of Houston, Houston, TX 77204, USA
}

\author{Micheal~Kahangirwe\,\orcidlink{0000-0001-9144-6240}}
\affiliation{
 Center for Nuclear Research, Department of Physics, Kent State University, Kent, OH 44243, USA
}

\author{Jamie~M.~Karthein\,\orcidlink{0000-0003-2041-5206}}
\affiliation{
 Cyclotron Institute, Texas A\&M University, College Station, TX 77843, USA
}
\affiliation{
 Department of Physics and Astronomy, Texas A\&M University, College Station, TX 77843, USA
}

\author{Jacquelyn Noronha-Hostler\,\orcidlink{0000-0003-3229-4958}}
\affiliation{
 The Grainger College of Engineering, Illinois Center for Advanced Studies of the Universe, Department of Physics, University of Illinois at Urbana-Champaign, Urbana, IL 61801, USA
}
\author{Claudia Ratti\,\orcidlink{0000-0002-8335-567X}}
\affiliation{
 Department of Physics, University of Houston, Houston, TX 77204, USA
}

\author{Mikhail Stephanov\,\orcidlink{0000-0002-8335-567X}}
\affiliation{
 Physics Department and Laboratory for Quantum Theory at the Extremes, University of Illinois at Chicago, Chicago, IL 60607, USA
}%

\date{\today}

\begin{abstract}
Inferring the possible existence and location of a critical point on the QCD phase diagram through experimental observations requires equations of state that include both lattice QCD constraints and critical behavior.
In this manuscript, we present the first four-dimensional equation of state in the $(T,\mu_B,\mu_Q,\mu_S)$ space combining the lattice 4D $T^\prime$-Expansion Scheme with a critical contribution mapped from the 3D Ising model. We extend the earlier two-dimensional construction, the so-called Ising 2D-TExS in the $(T, \mu_B)$ plane, promoting the critical point to a parametrizable critical surface in the full four-dimensional thermodynamic space. 
Our framework provides a family of equations of state to systematically study the effects of criticality in realistic hydrodynamic simulations with $BQS$ conserved charges, going beyond the case of strangeness neutrality, by varying the geometry of the critical surface and the strength of criticality at finite $\mu_B$, $\mu_Q$ and $\mu_S$ separately.
\end{abstract}

\maketitle

\section{Introduction}
\label{sec:introduction}

The phase diagram of strongly interacting matter is a central object of study in nuclear physics, connecting heavy-ion collisions (HICs), the interior of neutron stars and their mergers, and the early Universe \cite{Stephanov:2004wx,Ratti:2018ksb,MUSES:2023hyz}.
At vanishing baryon density, lattice QCD has established that the transition from hadronic matter to the quark-gluon plasma is an analytic crossover \cite{Aoki:2006we}, with pseudo-critical temperature $T_0 = 158.0 \pm 0.6$~MeV \cite{Borsanyi:2020fev}.
Furthermore, the equation of state (EoS) in this case is known with high precision \cite{Borsanyi:2013bia,HotQCD:2014kol}.
However, at finite baryon chemical potential $\mu_B$, direct simulations are hindered by the fermion sign problem, and knowledge of the EoS relies on extrapolations from $\mu_B=0$ or imaginary $\mu_B$: Taylor expansions \cite{Allton:2005gk,Borsanyi:2012cr,Bazavov:2017dus}, analytical continuation \cite{deForcrand:2002hgr,DElia:2002tig,Bellwied:2015rza}, Pad\'e approximants \cite{Bollweg:2022rps,Clarke:2024seq}, and, more recently, the constant entropy contour expansion scheme \cite{Shah:2024img,Shah:2026pue} and  the $T^\prime$-Expansion Scheme (TExS) \cite{Borsanyi:2021sxv,Borsanyi:2022qlh}.
The latter consists of resumming the leading $\mu_B$-dependence into a shift of the temperature, and extends the reach of lattice-based extrapolations to $\mu_B/T \lesssim 3.5-4$ without the unphysical behavior that afflicts truncated Taylor series at large density. This scheme has most recently been applied to magnetized QCD matter \cite{Borsanyi:2026yna}.
Several QCD-inspired models and functional approaches predict that the crossover turns into a first-order transition at large $\mu_B$, terminating at a critical point (CP) \cite{Halasz:1998qr,Berges:1998rc,Stephanov:1998dy,Stephanov:1999zu,Fu:2019hdw,Fischer:2018sdj}.

While a direct determination of the existence or location of the CP from first-principle QCD is as-yet unfeasible, the scaling properties near the critical point are known and are expected to belong to the universality class of the three-dimensional Ising model \cite{Rajagopal:1992qz,Berges:1998rc}.
Estimates on the location of the CP have narrowed down the acceptable region with agreement across various approaches \cite{Zhang:2026dny}.
Recent lattice QCD results exclude the critical point for $\mu_B < 450$ MeV \cite{Borsanyi:2025dyp} with complementary estimates from Lee-Yang edge singularity analyses \cite{Basar:2023nkp,Clarke:2024seq,Adam:2025phc}.
The constant-entropy-contour method of Ref. \cite{Shah:2024img} yields $(T_c,\mu_{B,c}) \sim (114, 602)$~MeV, in agreement with functional QCD \cite{Fu:2019hdw,Gunkel:2021oya} and holographic \cite{Hippert:2023bel} predictions, which cluster around $\mu_B \sim 400$--$650$~MeV.
On the experimental side, the Beam Energy Scan II (BES-II) program at RHIC has completed its data taking, and the STAR Collaboration has reported precision measurements of net-proton cumulants at $\sqrt{s_{NN}} = 3.0$--$27$~GeV, reaching up to $\mu_B \sim 750$~MeV \cite{STAR:2025zdq}.
As emphasized in Ref.~\cite{STAR:2025zdq}, dynamical model calculations including a critical point are needed to interpret these data.
The Beam Energy Scan Theory (BEST) topical collaboration developed many theoretical tools needed for such simulations \cite{An:2021wof} (for recent reviews of the dynamical simulations of fluctuation observables relevant to the critical point search, see, e.g., Refs.~\cite{Du:2024wjm,Stephanov:2024mdj,Koch:2025cog}). 
The essential input to such simulations is an EoS that incorporates known universal properties of the critical point singularity as well as the data from first-principle lattice simulations at zero chemical potential.

The QCD phase diagram is customarily drawn in the $(T,\mu_B)$ plane, but it is in fact four-dimensional: each of the three conserved charges of QCD---baryon number $B$, electric charge $Q$, and strangeness $S$---carries its own chemical potential.
In ultra-relativistic HICs, global strangeness neutrality $\langle n_S\rangle=0$ and the charge-to-baryon ratio of the colliding nuclei $\langle n_Q\rangle = 0.4\,\langle n_B\rangle$, single out one trajectory in the $(T,\mu_B,\mu_Q,\mu_S)$ space, while neutron-star matter in weak equilibrium and the early Universe follow different ones.
Inhomogeneity of the HIC fireball also means that different regions (or fluid cells) in the same fireball could explore a broad region of the phase diagram around the average trajectory \cite{Plumberg:2024leb}. Hydrodynamic codes that evolve all three charges are now available \cite{Plumberg:2024leb,Monnai:2019hkn}. 
Lattice-based 4D equations of state at finite 
$(T, \mu_B, \mu_Q, \mu_S)$
have accordingly been constructed from Taylor expansions \cite{Noronha-Hostler:2019ayj,Monnai:2019hkn,Monnai:2024pvy} and, most recently, from a generalization of the TExS to an arbitrary direction in chemical-potential space (4D-TExS) \cite{Abuali:2025tbd}, built on a new set of continuum-estimated second- and fourth-order $BQS$ susceptibilities \cite{jahan_2025_16749145}.
However, these constructions do not yet incorporate a critical point singularity.


The first equation of state built both to reproduce first-principles lattice results where they are available and to contain a critical singularity of the correct universality class at an adjustable location was constructed by the BEST Collaboration \cite{Parotto:2018pwx,An:2021wof}.
The BEST EoS maps the parametric 3D Ising scaling EoS \cite{Schofield:1969zza,Guida:1996ep,Nonaka:2004pg} onto the $(T,\mu_B)$ plane through a linear, non-universal change of variables and matches the result to the lattice Taylor coefficients at $\mu_B=0$ up to $\mathcal{O}(\mu_B^4)$.
The construction was later extended to strangeness-neutral matter \cite{Karthein:2021nxe}, to include the metastable phase \cite{Karthein:2024zvs}, it was used to study higher-order cumulants \cite{Mroczek:2020rpm,Karthein:2025hvl}, and its region of thermodynamic stability and causality was mapped out \cite{Mroczek:2022oga}.
Because it inherits the truncated Taylor expansion, the BEST EoS is limited to $\mu_B \lesssim 450$~MeV. 
In Ref. \cite{Kahangirwe:2024cny}, this limitation was overcome by embedding the Ising singularity into the effective temperature $T^\prime$ of the TExS EoS, with a quadratic mapping in $\mu_B$ that automatically respects charge-conjugation symmetry and follows the curvature of the transition line, extending the coverage to $\mu_B \simeq 700$~MeV.
Alternative constructions that superimpose a scaling function on a crossover EoS have also been proposed \cite{Kapusta:2022pny}, and the constant-entropy-contour method has been used to build an EoS whose critical point is inherited from the lattice input rather than introduced by hand \cite{Shah:2026pue}.
All of these, however, live in the $(T,\mu_B)$ plane or along a single trajectory through the multi-chemical-potential space.

Extending the critical point to four dimensions is not a simple matter of bookkeeping.
Universality fixes the singular behavior near the CP but offers no guidance on how the critical region deforms as $\mu_Q$ and $\mu_S$ are varied.
With three conserved charges, the critical points trace out a two-dimensional critical surface in $(T,\mu_B,\mu_Q,\mu_S)$.
Effective models as well as functional calculations at strangeness neutrality show that the CP location depends on the charge and strangeness content of the medium \cite{Costa:2020dgc,Aryal:2021ojz,Ueda:2013sia,Fu:2026qnl}.
Most recently, the constant-entropy-contour method was applied across all directions in chemical-potential space, providing the first mapping of the critical surface anchored on lattice input \cite{Shah:2026mzu}.
These calculations have shown that the surface is approximately elliptic in the $(\mu_B,\mu_S)$ plane and hyperbolic in the $(\mu_B,\mu_Q)$ plane with its shape controlled by the off-diagonal susceptibilities $\chi_{11}^{BS}$ and $\chi_{11}^{BQ}$.
Furthermore, along the strangeness-neutral direction $\mu_{B,c}$ was found to increase by $40$--$100$~MeV relative to the pure-$\mu_B$ direction with $T_c$ essentially unchanged, while no critical point is found along the pure $\mu_Q$ or $\mu_S$ axes. 
These results provide concrete guidance on the geometry that a phenomenological parametrization of the critical surface should be able to accommodate.

In this work we present the first family of QCD equations of state containing a critical point in the 3D Ising universality class in the full four-dimensional space of temperature and $BQS$ chemical potentials.
Building on the 4D-TExS framework \cite{Abuali:2025tbd}, we use a spherical decomposition of the chemical-potential vector, so that along each direction $(\theta,\varphi)$ the problem reduces to a one-dimensional extrapolation in the magnitude $\hmu$, and the Ising scaling variables $(r,h)$ can be mapped onto $(T',\hmu)$ as in Ref. \cite{Kahangirwe:2024cny}.
The critical chemical potential in every direction is prescribed at once by a quadric surface in $(\mu_B,\mu_Q,\mu_S)$ with six parameters, and the critical temperature then follows from the transition curve of the 4D-TExS EoS in that direction.
The singular contribution enters through the effective temperature $T'$, and a subtraction of its low-order expansion guarantees that the lattice susceptibilities of order two and four are reproduced exactly at $\mu=0$, such that the resulting EoS differs from the non-critical 4D-TExS only at $\mathcal{O}(\hmu^5)$. 
At last, we discuss the decomposition of $T^\prime$ into lattice and critical contributions, and present results of different thermodynamic quantities along the strangeness-neutral direction relevant for HICs. 
The code generating this family of equations of state will be made publicly available upon publication, and will be integrated into the MUSES calculation engine \cite{ReinkePelicer:2025vuh, Jahan:2026hvs}. 

\section{4D-TE\lowercase{x}S}
\label{sec:4dtexs}

The $T^\prime$-Expansion Scheme reorganizes the Taylor expansion of the lattice QCD EoS, by absorbing the leading dependence on the chemical potentials into a shift of the temperature at which a zero-density observable is evaluated~\cite{Borsanyi:2021sxv,Borsanyi:2022qlh}. Its four-dimensional extension treats temperature together with the three chemical potentials associated with baryon number $B$, electric charge $Q$, and strangeness $S$ \cite{Abuali:2025tbd}. 
Here, we use the dimensionless chemical potentials $\hmu_X \equiv \mu_X/T$, with $X\in\{B,Q,S\}$.
Following Ref.~\cite{Abuali:2025tbd}, we introduce the generalized chemical potential $\hmu$ and express the three conserved-charge chemical potentials in spherical coordinates as:
\begin{multline}
    \bm\hmu\equiv(\hmu_B,\hmu_Q,\hmu_S)\equiv\hmu\bm u; 
    \\[2pt]
    \bm u =(\cos\theta, \sin\theta \cos\varphi, \sin\theta \sin\varphi) \
\label{eq:texs_spherical_coordinates}
\end{multline}
where the 3 coordinates of $\bm u$ are referred to as $u_B$, $u_Q$ and $u_S$ respectively, based on the corresponding $\mu_X$ they map to. 
The vector $\bm u$ thus defines a direction in the chemical-potential space fixed by the angles $(\theta,\varphi)$, which are defined by:
\begin{equation}
\begin{aligned}
    \hmu &= \sqrt{\hmu_B^2+\hmu_Q^2+\hmu_S^2} \,,\\
    \theta &= \arccos\left(\hmu_B/\hmu\right) \,,\\
    \varphi &=
    \begin{cases}
        0 & \text{if }\hmu_Q^2 + \hmu_S^2 = 0, \\
        \arctan\!2(\hmu_S,\hmu_Q) & \text{otherwise}  \,,
    \end{cases}
\end{aligned}
\label{eq:texs_inverse_angles}
\end{equation}
with $\arctan\!2(x,y)$ the four-quadrant inverse tangent function taking values in $(-\pi, \pi]$.
This change to spherical coordinates allows the one-dimensional TExS from Refs.~\cite{Borsanyi:2021sxv,Borsanyi:2022qlh} to be applied along any arbitrary direction in the three-dimensional chemical-potential space~\cite{Abuali:2025tbd}. At fixed angles $(\theta,\varphi)$, the thermodynamic variables are $(T,\hmu)$, and the extrapolation is performed in the generalized chemical potential $\hmu$ along the specified direction defined by the corresponding $\bm u$ vector.

\subsection{Generalized susceptibilities}
\label{subsec:generalized_susceptibilities}
Let
\begin{equation}
    \chi_{ijk}^{BQS}(T)
    \equiv
    \left.
    \frac{\partial^{i+j+k}(P/T^4)}
    {\partial\hmu_B^i\,\partial\hmu_Q^j\,\partial\hmu_S^k}
    \right|_{{\hmu}=0}
\label{eq:bqs_susceptibilities}
\end{equation}
denote the susceptibilities calculated by lattice QCD at vanishing chemical potential \cite{Karsch:2010ck, Borsanyi:2011sw}.
At fixed $T$, $\theta$ and $\varphi$, the derivative w.r.t. the generalized chemical potential $\hmu$ is written from the chain rule as~\cite{Abuali:2025tbd}:
\begin{equation}
    \frac{\partial}{\partial\hmu}
    = \frac{\partial\hmu_B}{\partial\hmu}
      \frac{\partial}{\partial\hmu_B}
    + \frac{\partial\hmu_Q}{\partial\hmu}
      \frac{\partial}{\partial\hmu_Q}
    + \frac{\partial\hmu_S}{\partial\hmu}
      \frac{\partial}{\partial\hmu_S}.
\label{eq:texs_radial_chain_rule}
\end{equation}
By applying this derivative to the dimensionless pressure $\hat P = P/T^4$, one obtains the generalized susceptibilities defined by: 
\begin{align}
    X_n^{\theta,\varphi}(T)
    &\equiv
    \left.\frac{\partial^n(\hat P^{\theta,\varphi})}{\partial\hmu^n}
    \right|_{\hmu=0} 
    \nonumber\\
    &= \sum_{\substack{i,j,k\geq0\\i+j+k=n}}
    \frac{n!}{i!j!k!}\,
    u_B^i u_Q^j u_S^k\,
    \chi_{ijk}^{BQS}(T).
\label{eq:directional_susceptibility}
\end{align}
Hence, the generalized second-order susceptibility in the $(T,\hmu)$ plane reads:
\begin{align}
    X_2^{\theta,\varphi}(T)
    =&\, {u_B}^2 \, \chi_2^B(T)
     + {u_Q}^2 \, \chi_2^Q(T)
     + {u_S}^2 \, \chi_2^S(T) \nonumber\\
    &+2 u_B u_Q \chi_{11}^{BQ}(T)
     +2 u_B u_S \chi_{11}^{BS}(T) \nonumber\\
    &+2 u_Q u_S \chi_{11}^{QS}(T).
\label{eq:X2_directional}
\end{align}
Similarly, one can derive the corresponding generalized fourth-order susceptibility $X_4^{\theta,\varphi}$ from Eq.~\eqref{eq:directional_susceptibility}, involving all
15 pure and mixed fourth-order $BQS$ susceptibilities separately.

\subsection{Stefan--Boltzmann-normalized expansion}
\label{subsec:texs_expansion}

Following the Stefan-Boltzmann (SB)-normalized prescription of Ref.~\cite{Borsanyi:2022qlh}, the central 4D-TExS relation expresses the generalized density $X_1$ at finite $\hmu$ in terms of the zero-density generalized second-order susceptibility $X_2$, evaluated at an effective temperature $T^\prime$ for a given $\bm u$ direction:
\begin{equation}
    X_1^{\theta,\varphi}(T,\hmu)=\frac{\overline{X}_1^{\theta,\varphi}(\hmu)}{\overline{X}_2^{\theta,\varphi}(0)}
    X_2^{\theta,\varphi}
    \!\left(T_{\rm lat}^{\prime\,\theta,\varphi}(T,\hmu),0\right).
\label{eq:4dtexs_definition}
\end{equation}
Here, the barred quantities denote the Stefan-Boltzmann (SB) limits, i.e., the value of the corresponding quantities when $T \to \infty$ and the system behaves as an ideal gas of massless quarks and gluons. The SB pre-factor also retains the correct high-temperature dependence on $\hmu$ in $\overline{X}_1^{\theta,\varphi}$, which is obtained by taking the $\hmu$ derivative of the Taylor-expanded pressure around $\hmu = 0$, then taking its limit when $T \to 0$:  
\begin{equation}
    \overline{X}_1^{\theta,\varphi}(\hmu)
    = \hmu \, \overline{X}_2^{\theta,\varphi}(0)
    + \frac{\hmu^3}{3!} \, \overline{X}_4^{\theta,\varphi}(0)
    \, ,
\end{equation}
and the SB limits $\overline{X}_2^{\theta,\varphi}$ and $\overline{X}_4^{\theta,\varphi}$ at $\hmu=0$ are simply calculated based on Eq.~\eqref{eq:directional_susceptibility}.

The regular part of the effective temperature, used to construct the pure-lattice expansion as defined in Ref.~\cite{Borsanyi:2022qlh} and denoted $T_{\rm lat}^{\prime\,\theta,\varphi}$ here, is expanded as:
\begin{equation}
 T_{\rm lat}^{\prime\,\theta,\varphi}(T,\hmu)
    = T\left[1+\lambda_2^{\theta,\varphi}(T)\hmu^2
    +\mathcal{O}(\hmu^4)\right]
    \, ,
    \label{eq:Tprime_lattice}
\end{equation}
with the leading expansion coefficient, derived in similar fashion to Ref.~\cite{Borsanyi:2021sxv}, being defined as:
\begin{equation}
    \begin{aligned}
    \lambda_2^{\theta,\varphi}(T)
    = \frac{1}{6T\,\partial_T X_2^{\theta,\varphi}(T)} \times
    & \\ & \hspace{-1.5cm}
    \biggl[
    X_4^{\theta,\varphi}(T)
    -\frac{\overline{X}_4^{\theta,\varphi}(0)}{\overline{X}_2^{\theta,\varphi}(0)}
     X_2^{\theta,\varphi}(T)
    \biggr].
    \end{aligned}
\end{equation}
Thus, the finite-density extrapolation uses lattice susceptibilities at ${\hmu}=0$, while its angular dependence retains all $BQS$ correlations contained in those susceptibilities.  

\subsection{Thermodynamic reconstruction}
\label{subsec:texs_thermodynamics}

We reconstruct all the thermodynamic functions from the generalized charge density $X_1^{\theta, \varphi}(T,\hmu)$, following the procedure used in the TExS framework~\cite{Borsanyi:2021sxv,Borsanyi:2022qlh}. Since by definition $X_1=\partial\hat P/\partial\hmu$ at fixed angles, the pressure follows as:
\begin{equation}
    \hat P(T,\hmu)
    = \hat P(T,0)
    + \int_0^{\hmu}
    X_1(T,x)\,dx.
\label{eq:texs_pressure}
\end{equation}
Note that we are voluntarily omitting the explicit ${\theta,\varphi}$ dependence of thermodynamic quantities from now on, to make the notation lighter.
The generalized charge density alone helps to determine the pressure by integration, but the individual conserved-charge densities $\hat n_X = n_X / T^3$ also require the angular variation of the pressure. Defining the integrals:
\begin{equation}
    I_\ell(T,\hmu)
    \equiv \int_0^{\hmu}\partial_\ell
    X_1(T,x)\,dx
    \qquad \text{with }\ell\in\{\theta,\varphi\},
\label{eq:angular_integrals}
\end{equation}
the spherical-coordinate chain rule gives:
\begin{equation}
    \hat{n}_X 
    = u_X X_1(T,\hmu) 
    + \frac{\partial\theta}{\partial\hmu_X} I_\theta
    + \frac{\partial\varphi}{\partial\hmu_X} I_\varphi ,
    \label{eq:component_densities}
\end{equation}
with detailed derivation given in the appendices of Ref.~\cite{Abuali:2025tbd}.
The apparent singularities at $\hmu=0$ and on the polar axis are understood through their continuous limits. Retaining the angular terms in Eq.~\eqref{eq:component_densities} is thus essential away from the baryon axis. 

Other thermodynamic observables follow from the reconstructed pressure. With $\hat\epsilon = \epsilon/T^4$ and $\hat s=s/T^3$, we have:
\begin{equation}
    \hat s(T,\hmu)
    =4\hat P + T\left.\frac{\partial\hat P}{\partial T}\right|_{\hmu,\theta,\varphi} - \hmu X_1,
\label{eq:texs_entropy}\\
\end{equation}
\begin{equation}
    \hat\epsilon(T,\hmu) =\hat s-\hat P+\hmu X_1.
\label{eq:texs_energy}
\end{equation}
Equations~\eqref{eq:4dtexs_definition}--\eqref{eq:texs_energy} define the non-critical 4D-TExS baseline.  The next section modifies its effective temperature so that the equation of state inherits the nonanalytic structure of the three-dimensional Ising universality class while preserving the lattice expansion at low chemical potential.

\section{Embedding the Critical Point}
\label{sec:critical-point}

Critical signatures are expected to become pronounced near the critical point, where the correlation length diverges and thermodynamic response is governed by universal scaling rather than microscopic details. Systems with the same symmetry and dimensionality therefore exhibit the same critical behavior. The QCD critical point, if present in the phase diagram, is expected to belong to the 3D-Ising universality class \cite{Rajagopal:1992qz}. This correspondence allows the 3D Ising model to be incorporated into the 4D-TExS equation of state introduced in Sec.~\ref{sec:4dtexs}. This construction proceeds in three steps:  
expressing the 3D-Ising EoS in terms of its scaling variables;
mapping Ising variables to the $(T,\hmu)$ plane at fixed $(\theta,\varphi)$; 
explore the critical surface in three conserved charges, while matching the critical contribution to the $T^{\prime\,\theta,\varphi}$ from  4D-TExS.

\subsection{3D-Ising scaling}
\label{subsec:Ising_scaling}

We start by reviewing the scaling of the singular part of the thermodynamic potential within the standard Josephson-Schofield parametric $(R,\Theta)$ representation \cite{Guida:1996ep,Schofield:1969zza,Nonaka:2004pg} of the 3D Ising model. Note that the use of $\Theta$ here is to avoid confusion with $\theta$ rotations in conserved charge chemical potential space, as defined in Sec. \ref{sec:4dtexs}. The construction is specified by the universal critical exponents
\begin{align}
    \alpha &= 0.11 \, , &
    \beta  &= 0.326 \, , &
    \delta &= 4.8 \,
    \label{eq:ising_exponents}
\end{align}
together with the auxiliary functions
\begin{align}
    \tilde h(\Theta) &= \Theta\bigl(1 + a\,\Theta^2 + b\,\Theta^4\bigr) \, , \\
    g(\Theta)        &= c_0 + c_1(1-\Theta^2) + c_2(1-\Theta^2)^2 + c_3(1-\Theta^2)^3 \, ,
\end{align}
with $a = -0.76201$, $b = 0.00804$, and entering $g(\Theta)$ are
\begin{equation}
\begin{aligned}
c_0 &= \frac{\beta}{2-\alpha}(1+a+b), \\
c_1 &= -\frac{1}{2(\alpha-1)}
\bigl[(1-2\beta)(1+a+b)-2\beta(a+2b)\bigr], \\
c_2 &= -\frac{1}{2\alpha}
\bigl[2\beta b-(1-2\beta)(a+2b)\bigr], \\
c_3 &= -\frac{b(1-2\beta)}{2(\alpha+1)}.
\label{eq.c's}
\end{aligned}
\end{equation}

In the 3D Ising model, the magnetization $M$ is the order parameter, while the reduced temperature $r$ and external field $h$ are the scaling variables. They are expressed in terms of $(R,\Theta)$ as:
\begin{equation}
    \begin{aligned}
    M(R,\Theta) &= M_0\,R^\beta\,\Theta \, , \\
    h(R,\Theta) &= h_0\,R^{\beta\delta}\,\tilde{h}(\Theta) \, , \\
    r(R,\Theta) &= R\left(1-\Theta^2\right) \, ,
    \end{aligned}
\end{equation}
where $M_0$ and $h_0$ are normalization constants:
\begin{equation}
    M_0 = 0.605,
    \qquad
    h_0 = 0.364.
\label{eq:normal-cste_Ising}
\end{equation}

The corresponding singular contribution to the Ising pressure is:
\begin{equation}
\begin{aligned}
P_\mathrm{Ising}(R,\Theta)
&=
h_0 M_0\,R^{2-\alpha}
\left[
\Theta\,\tilde{h}(\Theta)-g(\Theta)
\right],
\end{aligned}
\end{equation}
where $\tilde h(\Theta)$ and $g(\Theta)$ encode the angular dependence of the Ising equation of state.

\subsection{Mapping the Ising critical region into 4D-TExS}
\label{subsec:ising-4dtexs-mapping}

\begin{figure*}[!thb]
    \centering
    \includegraphics[width=\textwidth]{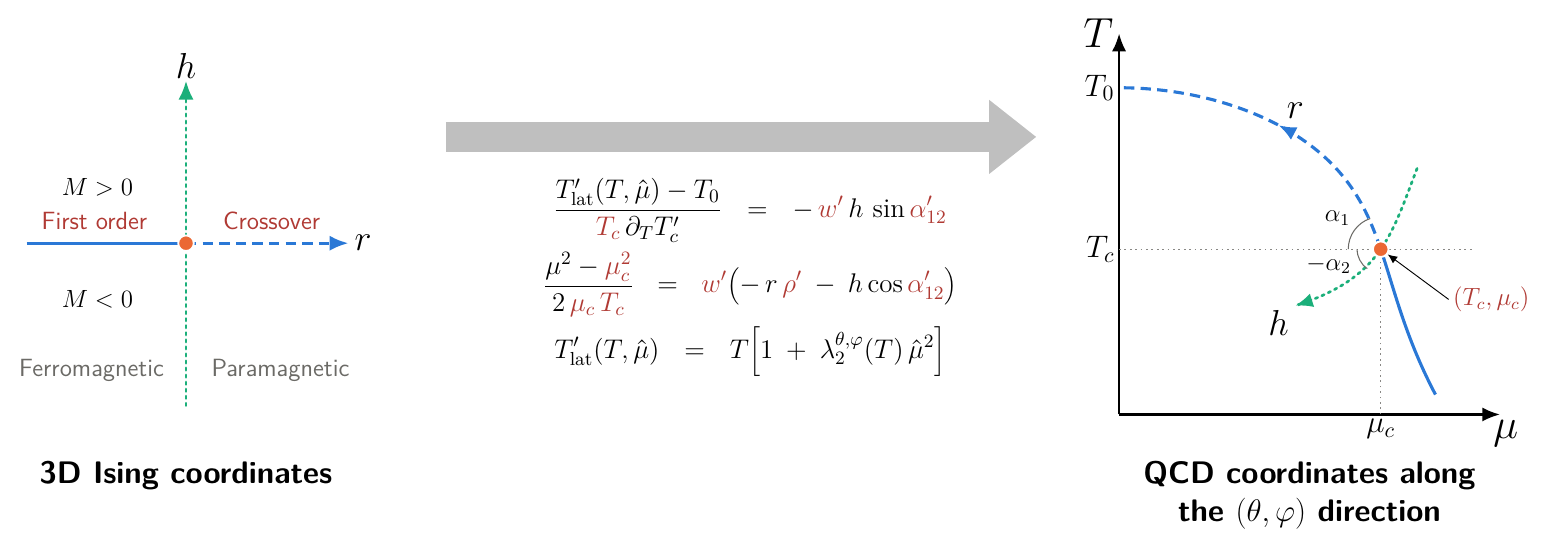}
    \caption{
        Schematic mapping of the 3D-Ising critical structure into the 4D-TExS thermodynamic space. Left: the Ising scaling plane $(r,h)$. On the line $h=0$, the critical point at $(r,h)=(0,0)$ separates the first-order branch ($r<0$, solid) from the crossover ($r>0$, dashed). Center: the mapping of Eq. \eqref{eq:mapping}. Non-universal quantities are shown in red. Right: the $(T,\mu)$ plane along the $(\theta,\varphi)$ direction. The image of the $r$ axis is the transition curve $T'_{\rm lat}(T,\hmu)=T_0$, crossover (dashed) above the critical point $(T_c,\mu_c)$ and first order (solid) below it. The dotted curve through the critical point is the image of the $h$ axis, and $\alpha_1$ and $-\alpha_2$ are the angles of the two images with the $\mu$ direction at the critical point. }
    \label{fig:mapping}
\end{figure*}

The 3D-Ising critical structure is mapped onto the 4D-TExS thermodynamic space through a non-universal transformation adapted from Ref.~\cite{Kahangirwe:2024cny}. The 4D-TExS equation of state depends on temperature and three conserved-charge chemical potentials $(T,\hmu_B,\hmu_Q,\hmu_S)$, whereas the 3D Ising model is described by only two scaling variables, $(r,h)$. 
To connect these spaces, we use the spherical decomposition introduced in Eq.~\eqref{eq:texs_spherical_coordinates}. 
For a fixed direction $(\theta,\varphi)$ in chemical-potential space, the angular coordinates remain fixed and only the radial generalized chemical-potential coordinate $\hmu$ varies. The mapping therefore reduces, along each direction $\bm u$, to a two-dimensional transformation between the Ising variables $(r,h)$ and the QCD variables $(T,\mu)$. Varying $(\theta,\varphi)$ then generates the corresponding mappings throughout the full conserved-charge chemical-potential space. 
Two central features of the present construction, first introduced in Ising 2D TExS \cite{Kahangirwe:2024cny} as in Fig.~\ref{fig:mapping}, are the quadratic nature of $\mu$ which captures charge conjunction symmetry, and the choice of fixing the Ising axis $r$ to follow the transition curve. This is different from the BEST linear mapping introduced in Ref.~\cite{Parotto:2018pwx}, where the Ising axes are introduced through a local mapping around the critical point. For notational simplicity, we suppress the explicit $(\theta,\varphi)$ dependence in what follows.

For a given vector $\bm u$, the mapping is defined as:
\begin{equation}
\begin{gathered}
    \frac{ T_{\rm lat}^{\prime}(T,\hmu)-T_0} {T_c\,\partial_T{T}_c^\prime}
    = -\,w^{\prime} \,h(T,\mu) \sin\alpha_{12}^{\prime},
    \\
    \frac{\mu^2-\mu_c^2}{2\mu_c T_c} 
    = w^{\prime} \left[ -\,\rho^{\prime} \,r(T,\mu) -\, h(T,\mu) \,\cos\alpha_{12}^{\prime} \right]
\end{gathered}
\label{eq:mapping}
\end{equation}
where $T_0$ is the pseudocritical temperature at vanishing chemical potentials and
$\left(T_c,\mu_c\right)$ specifies the location of the critical point along the selected direction. 
We further define
$\partial_T T_c^\prime\equiv\left.\left(\frac{\partial T^\prime}{\partial T}\right)_{\hmu}\right|_{T=T_c,\,\hmu=\hmu_c}.$ 

The parameters $w^{\prime}$, $\rho^{\prime}$, and $\alpha_{12}^{\prime}$ are non-universal and effectively affect the critical region properties:
$w^\prime$ sets the overall size of the critical region, $\rho^\prime$ controls the relative scaling of the two Ising directions, and $\alpha_{12}^\prime$ is an effective angle, not the geometric angle between two physical axis $(T,\mu)$ but it contributes in fixing  the effective orientation of the scaling axes in the transformed coordinates. Because these parameters are not fixed by universality, they must be specified through additional non-universal input. Following Ref.~\cite{Kahangirwe:2024cny}, we relate the quadratic mapping in Eq.~\eqref{eq:mapping} to the conventional linear mapping of Ref.~\cite{Parotto:2018pwx}. The latter is characterized by the angle $\alpha_1$, which specifies the slope of the transition line at the critical point, the auxiliary angle $\alpha_2=\alpha_1-\alpha_{12}$, and the scale parameters $w$ and $\rho$. 
Expanding the quadratic mapping about the critical point and matching the coefficients multiplying the Ising variables $(r,h)$ to those of the mapping yields:

\begin{equation}
\begin{aligned}
    & \tan\alpha_{12}' = \tan\alpha_1-\tan\alpha_2,
    \\
    & w^{\prime} = \frac{w}{\cos\alpha_1}\sqrt{(\cos\alpha_1\cos\alpha_2)^2+\sin^2 \alpha_{12}},
    \\
    & \rho^{\prime} = \rho\,\frac{\cos^2\alpha_1}{\sqrt{(\cos\alpha_1\cos\alpha_2)^2+\sin^2\alpha_{12}}} \, .
\end{aligned}
\label{eq:linear-relations}
\end{equation}

Thus, the primed parameters play the same physical roles as their counterparts in the linear mapping, but are related to them through the geometric transformation in Eq.~\eqref{eq:linear-relations}. In particular, the direction dependence enters through $\alpha_1=\alpha_1(\theta,\varphi)$, which is determined by the local slope of the transition line at the critical point. Consequently, $w^\prime$, $\rho^\prime$, and $\alpha_{12}^\prime$ also inherit an implicit dependence on $(\theta,\varphi)$.

\subsection{Transition line and critical point location}
\label{subsec:mapping_geometry}

The transition line is naturally defined in Ising coordinates by $h=0$, which translates to QCD coordinates into the condition given by Eq.~\eqref{eq:CP_condition}. The transition line is chosen such that, along any $\bm u$, it follows the $T^\prime$-expansion of the lattice QCD pseudocritical temperature $T_{\rm lat}^{\prime}(T_c,\hmu)$ and reduces at vanishing chemical potentials to $T_0$, the pseudocritical temperature of the chiral crossover \cite{Borsanyi:2020fev}:
\begin{equation}
    T_{\rm lat}^{\prime}(T_c,\hmu) = T_c\left[ 1+\lambda_2(T_c) \,\hmu^2 \right] = T_0 .
\label{eq:CP_condition}
\end{equation}

For a specified critical chemical potential $\mu_c$, Eq.~\eqref{eq:CP_condition} determines the corresponding critical temperature $T_c$. The critical point $(T_c,\mu_c)$ therefore lies on the same transition curve that separates the crossover and first-order branches in the mapped equation of state. The slope of the transition curve at the critical point determines the angle $\alpha_1$, giving (as derived in Appendix~\ref{apdx:transition_slope}):
\begin{equation}
    \tan\alpha_1 = \left| \left. \frac{dT_c}{d\mu} \right|_{T_c, \mu_c}\right| 
    =
   \left|  -\frac{
    2\,\lambda_2(T_c)\,\hmu_c
    }{
    1+{\hmu_c}^2
    \left[
    T_c\lambda_2^{\prime}(T_c)-\lambda_2(T_c)
    \right]} \right| .
\label{eq:slope}
\end{equation}
Equation~\eqref{eq:slope} therefore fixes $\alpha_1$ directly from the local slope of the 4D-TExS transition curve at the critical point. Through Eq.~\eqref{eq:linear-relations}, this angle determines the corresponding primed mapping parameters. Once $\alpha_1$ is fixed in this way, $\alpha_2$ or $\alpha_{12}$ remains the only independent angular parameter controlling the orientation of the Ising scaling axes relative to the local 4D-TExS transition curve.

\subsection{Matching criticality in the \texorpdfstring{$T^\prime$-expansion}{}}
\label{subsec:matching_CP}

Within the 4D-TExS framework, the finite-density dependence of the equation of state is encoded through the effective temperature variable \(T^{\prime}\), defined in Eq.~\eqref{eq:Tprime_lattice}. We introduce critical behavior along a fixed direction \((\theta,\varphi)\) in the  generalized conserved-charge space, by modifying the effective temperature, separating the regular contribution from the singular one associated with the 3D-Ising universality class. We define
\begin{align}
    T_{\rm full}^{\prime}(T,\hmu)
    =&
    T_{\rm lat}^{\prime}(T,\hmu)
    \nonumber \\ 
    & \quad
    \underbrace{
    +\, T_{\rm Ising}^{\prime}(T,\hmu)
    - T_{\rm Taylor}^{\prime}(T,\hmu)
    }_{T_{\rm crit}^{\prime}(T,\hmu)}
\label{eq:Tfull}
\end{align}
where $T_{\rm lat}^{\prime}$ denotes the non-critical contribution constrained by the lattice-QCD expansion, while $T_{\rm crit}^{\prime}$ introduces the desired critical structure.

The critical contribution is constructed from the mapped 3D Ising density $n_{\rm Ising}=\partial P_{\rm Ising}/{\partial \hmu}$ as
\begin{equation}
    T_{\rm Ising}^{\prime}
    =
    \left( \left.
    \frac{\partial X_{2,{\rm lat}}}{\partial T}
    \right|_{T_0} \right)^{-1}
    \frac{n_{\rm Ising}(T,\hmu)}
    {\overline{X}_1(\hmu)/\overline{X}_2(0)},
\label{eq:Tcrit-critical-point}
\end{equation}
so that the nonanalytic behavior generated by the Ising EoS is transferred directly to $T^{\prime}_{\rm Ising}$, which is derived explicitly in Appendix~\ref{apdx:Ising_density}.
The subtracted $T_{\rm Taylor}^{\prime}$ term in Eq.~\eqref{eq:Tfull} prevents double counting of low-order Taylor contributions already contained in the $T_{\rm lat}^{\prime}$, and preserves the lattice expansion around \(\hmu=0\) to the same order used in constructing the regular 4D-TExS equation of state. 
Explicitly:
\begin{align}
    T_{\rm Taylor}^{\prime}(T,\hmu)=
    &\left(\left.\frac{\partial X_{2,{\rm lat}}}{\partial T}\right|_{T_0}\right)^{-1}\\
    &\times \frac{\left[\left.\frac{\partial n_{\rm Ising}}{\partial\hmu}\right|_{\hmu=0}\hmu + \frac{1}{3!} \left.\frac{\partial^3 n_{\rm Ising}}{\partial\hmu^3}\right|_{\hmu=0}\hmu^3
    \right]
    }{\overline{X}_1(\hmu)/\overline{X}_2(0)}.
\label{eq:T-taylor}
\end{align}
This form follows from the small-$\hmu$ expansion of the numerator of Eq. \eqref{eq:Tcrit-critical-point}. The Ising density at fixed $(T,\theta,\varphi)$ is odd in $\hmu$. Its expansion
\begin{equation}
n_{\rm Ising}(T,\hmu)
=X_{2,{\rm Ising}}(T)\,\hmu
+\frac{X_{4,{\rm Ising}}(T)}{3!}\,\hmu^3
+\mathcal{O}(\hmu^5),
\label{eq:nIsing_taylor}
\end{equation}
defines $X_{2,{\rm Ising}}=\partial n_{\rm Ising}/\partial \hmu|_{\hmu=0}$ and $X_{4,{\rm Ising}}=\partial^3 n_{\rm Ising}/\partial \hmu^3|_{\hmu=0}$, the second and fourth-order susceptibilities of the Ising pressure. 
The SB pressure is a polynomial of fourth degree in $\hmu$, therefore it is given exactly as:
\begin{equation}
\frac{\overline{X}_1(\hmu)}{\overline{X}_2(0)}
=\hmu\left(1+\bar{\kappa}\,\frac{\hmu^2}{3!}\right),
\qquad
\bar{\kappa}\equiv\frac{\overline{X}_4(0)}{\overline{X}_2(0)}.
\label{eq:SB_prefactor}
\end{equation}

\begin{figure*}[!ht]
    \centering
    \includegraphics[width=\textwidth]{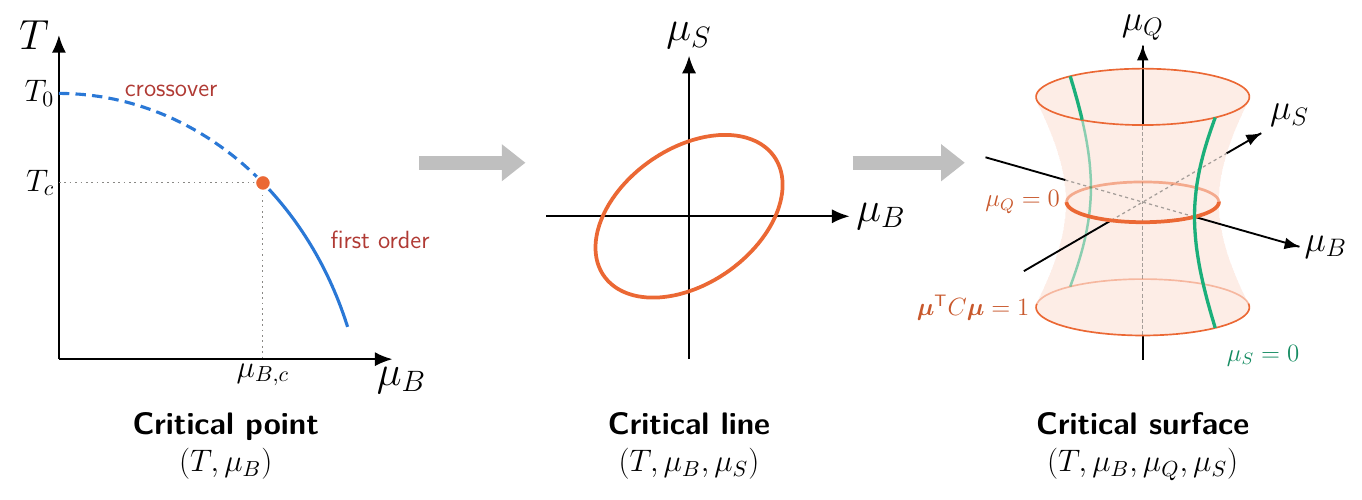}
    \caption{Schematic critical structure as conserved charges are added. Left: with one charge, the critical point $(T_c,\mu_B^c)$ lies on the transition line in the $(T,\mu_B)$ plane and separates its crossover and first-order branches. Center: with two charges, the critical points form a line in $(T,\mu_B,\mu_S)$ that, under the quadratic ansatz, has a projection on the chemical potentials plane which is a conic, here a tilted ellipse. Right: with three charges, the chemical-potential projection is prescribed by $\bm{\mu}^{\mathsf T}C\,\bm{\mu}=1$ in Eq. \eqref{eq:critical_surface}, here a one-sheet hyperboloid.}
    \label{fig:critical_point_surface}
\end{figure*}

The leading power of $\hmu$ cancels in Eq. \eqref{eq:Tcrit-critical-point}, which becomes
\begin{equation}
    \begin{aligned}
    T_{\rm Ising}^{\prime}=
    &\left(\left.\frac{\partial X_{2,{\rm lat}}}{\partial T}\right|_{T_0}\right)^{-1}\\
    &\times\frac{X_{2,{\rm Ising}}+X_{4,{\rm Ising}}\,(\hmu^2/3!)+\mathcal{O}(\hmu^4)}
    {1+\bar{\kappa}\,(\hmu^2/3!)}.
    \end{aligned}
\label{eq:TIsing_expansion}
\end{equation}
On its own, $T_{\rm Ising}^{\prime}$ would therefore shift $T^\prime$ at $\hmu=0$ by $\bigl(\partial X_{2,{\rm lat}}/\partial T|_{T_0}\bigr)^{-1}X_{2,{\rm Ising}}$. The terms $X_{2,{\rm Ising}}\,\hmu$ and $X_{4,{\rm Ising}}\,(\hmu^3/3!)$ of $n_{\rm Ising}$ are the Ising contributions to the susceptibilities $X_2$ and $X_4$, which the lattice data already contain and from which $\lambda_2$ is determined.
Therefore, they would be counted twice and would change the coefficients $X_2$ and $X_4$ of $X_1$. Subtracting Eq. \eqref{eq:T-taylor} removes exactly these two terms from the Ising density. The SB factor is exact and common to $T_{\rm Ising}^{\prime}$ and $T_{\rm Taylor}^{\prime}$, so it is not expanded:
\begin{equation}
    \begin{aligned}
        T^\prime_{\rm crit}&=T_{\rm Ising}^\prime - T_{\rm Taylor}^\prime\\
        &=\left(\left.\frac{\partial X_{2,{\rm lat}}}{\partial T}\right|_{T_0}\right)^{-1}\\
        & \qquad \times\frac{n_{\rm Ising}-X_{2,{\rm Ising}}\,\hmu-X_{4,{\rm Ising}}\,(\hmu^3/3!)}{\hmu\left[1+\bar{\kappa}\,(\hmu^2/3!)\right]}.
    \end{aligned}
\end{equation}
Hence, $X_1$ differs from the noncritical 4D-TExS density only at $\mathcal{O}(\hmu^5)$: its coefficients $X_2$ and $X_4$ are the lattice ones. Since the numerator and the SB term of Eq. \eqref{eq:T-taylor} grow with the same power of $\hmu$, $T_{\rm Taylor}^{\prime}$ stays bounded at large $\hmu$, consistently with the expansion.

Replacing \(T_{\rm lat}^{\prime}\) by
\(T_{\rm full}^{\prime}\) in Eq.~\eqref{eq:4dtexs_definition}
therefore incorporates the critical contribution into the 4D-TExS construction, while retaining consistency with the lattice QCD expansion at small \(\hmu\). 
In particular, the resulting conserved-charge density acquires the required nonanalytic critical structure near the mapped critical point, whereas its low-density behavior remains matched to the lattice QCD expansion baseline.

\subsection{Parameterizing the critical surface}
\label{subsec:critical_surface}

The mapping in Eq.~\eqref{eq:mapping} acts in one direction at a time along each direction $\bm u$ and requires the critical-point location $(T_c,\mu_c)$. Rather than choosing these points independently for every direction, we prescribe the critical chemical potential of all directions at once through a single surface in the space of the three chemical potentials. 
This follows from the fact that, with one conserved charge, the critical point lies on the transition line; with two charges, the critical points form a line; and with three charges, they form a surface. Fig.~\ref{fig:critical_point_surface} illustrates this progression. The critical temperature of each direction then follows from its transition curve, as defined in Eq.~\eqref{eq:CP_condition}, whenever an admissible positive solution exists. 

The geometry of this critical structure contains physical information about how the critical point moves as additional chemical potentials are varied. Effective-model studies have demonstrated that its location depends on the charge and strangeness content of matter \cite{Costa:2020dgc,Aryal:2021ojz,Ueda:2013sia}. 
More directly, a critical surface in the full $(T,\mu_B,\mu_Q,\mu_S)$ space has been mapped using a second-order expansion along contours of constant entropy density, anchored on lattice QCD inputs \cite{Shah:2026mzu}. Sections in the $(\mu_B,\mu_S)$ plane have been found to be approximately elliptic, while hyperbolic in the $(\mu_B,\mu_Q)$ plane, which motivates a parameterization that can represent different geometries in different charge planes.

\begin{figure*}[ht]
    \centering
    \includegraphics[width=\textwidth]{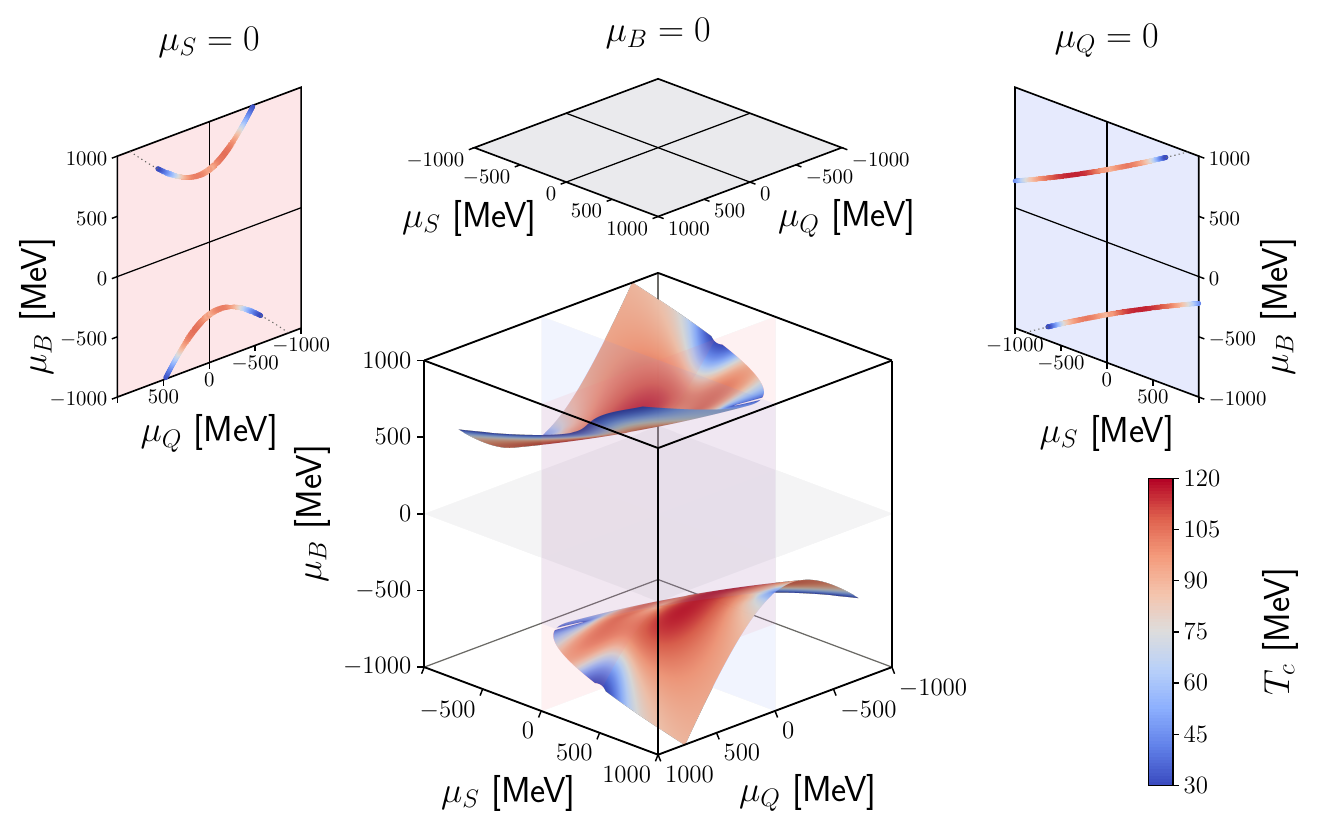}
    \caption{
    Example of a critical surface, parametrized according to Eq.~\eqref{eq:critical_surface}, with parameter values chosen as  $(C_{BB}, C_{BQ}, C_{BS}, C_{QQ}, C_{QS}, C_{SS}) = (2.76, 0.80, -1.15, -4.22, -0.40, 0.35)$
    GeV$^{-2}$.
    Center: the surface is colored by the value of $T_c$ obtained from Eq.~\eqref{eq:CP_condition}, with $T_0=158$ MeV and the average of $\lambda_2(T)$. 
    The shaded planes are $\mu_B=0$ (gray), $\mu_Q=0$ (blue), and $\mu_S=0$ (pink). Top: sections of the surface by the planes $\mu_S=0$, $\mu_B=0$, and $\mu_Q=0$, each drawn as it appears in the central view, with the same color scale. Dotted lines mark the parts of the section with a $T_c$ outside the EoS validity range. 
    }
    \label{fig:critical_surface_param}
\end{figure*}

As an illustrative example, we chose to describe the critical surface as a quadric. 
Charge-conjugation invariance of QCD, $p(T,\bm{\mu})=p(T,-\bm{\mu})$, requires the set of critical points to be symmetric under $\bm{\mu}\to-\bm{\mu}$, therefore the quadratic surface has to be centered at the origin. Its equation reads:
\begin{equation}
    \bm{\mu}^{\mathsf T}C\,\bm{\mu}=1,
    \qquad
    C=\begin{pmatrix}
        C_{BB} & C_{BQ} & C_{BS}\\
        C_{BQ} & C_{QQ} & C_{QS}\\
        C_{BS} & C_{QS} & C_{SS}
    \end{pmatrix}.
\label{eq:critical_surface}
\end{equation}

The quadratic form is the simplest choice consistent with charge-conjugation symmetry. The constant term must be nonzero because the surface cannot pass through $\bm{\mu}=0$, where the transition is a crossover. With this symmetry alone, fixing the normalization leaves six independent entries in the symmetric matrix $C$ having the units of inverse energy squared. The critical temperatures determined below complete the description of a two-dimensional surface in $(T,\mu_B,\mu_Q,\mu_S)$. Its projection on the chemical potential space contains only those points of Eq. \eqref{eq:critical_surface} that admit a physical critical temperature.

Along $\bm u$, Eq. \eqref{eq:critical_surface} reads $\mu^2\,\bm{u}^{\mathsf T}C\,\bm{u}=1$, so the critical chemical potential reads:
\begin{equation}
    \mu_c(\theta,\varphi)=\frac{1}{\sqrt{C_{\bm u\bm u}}},
    \qquad
    C_{\bm u\bm u}\equiv\bm{u}^{\mathsf T}C\,\bm{u}.
\label{eq:muc_direction}
\end{equation}
Directions with $C_{\bm u\bm u}\le0$ carry no critical point and along them the equation of state is the noncritical 4D-TExS. A geometric intersection having $C_{\bm u\bm u}>0$ is necessary but not sufficient: the transition curve must also admit the corresponding critical temperature within the domain of the equation of state.

For convenience, one can introduce $\mu_X^*$, which corresponds to the value of chemical potentials where the critical surface intersects with the $\mu_X$ axis, such that $\mu_X^* = 1/\sqrt{C_{XX}}$ according to Eq.~\eqref{eq:muc_direction}.

\section{Results}
\label{sec:results}

In this Section, we present our first thermodynamic results obtained from the EoS construction introduced in Sec.~\ref{sec:critical-point}. In order to visualize our four-dimensional EoS, we elect to display results obtained along the trajectory corresponding to the conditions satisfied on average by the system created in HICs, imposed by strangeness neutrality and $Z/A$ ratio of the colliding nuclei. For simplicity, we will refer to this scenario as ``average HIC conditions".
While it would be convenient to use the approximation of $\mu_S \approx \mu_B / 3$ and $\mu_Q \approx - \mu_B / 30$ \cite{Lysenko:2024hqp}, which corresponds to carry an expansion along a straight direction with $(\theta = 18.48^\circ, \varphi = 94.29^\circ)$ in our construction, we do show results corresponding to the exact average HIC condition trajectory, namely $\langle n_S \rangle = 0$ and $\langle n_Q/n_B \rangle = 0.4$.
Note that, since our EoS is built from grand-canonical lattice QCD susceptibilities, these conditions correspond, for each fixed values of $(T, \mu_B)$, to a unique set of finite $(\mu_Q, \mu_S)$ if such a set exists. Although the choice of displaying results along this trajectory is motivated by the fact that phenomenological modeling of HICs is expected to remain reasonably close to these average conditions, the Ising 4D-TExS EoS itself is not restricted to this trajectory and can describe any configuration throughout its domain of validity in $(T,\mu_B,\mu_Q,\mu_S)$.

The results presented hereafter rely on an EoS computed with the newest set of second- and fourth-order continuum-estimated $BQS$ lattice susceptibilities, obtained from the Wuppertal-Budapest Collaboration data \cite{Abuali:2025tbd,jahan_2025_16749145}.
For simplicity, the critical surface is parametrized as a sphere in the $(\mu_B,\mu_Q,\mu_S)$ space with $\mu_B^* = \mu_Q^* = \mu_S^* = 520$ MeV, corresponding to $C_{BB} = C_{QQ} = C_{SS} = 3.7$ GeV$^{-2}$ as the only non-zero parameters in Eq.~\eqref{eq:critical_surface}.
Along the average HIC condition trajectory, this parameterization gives thus a critical point at $(T, \mu_{B}, \mu_{Q}, \mu_{S})_c = (142, 492, -34, 164)$ MeV.

For the 3D-Ising-to-QCD mapping parameters introduced in Sec.~\ref{subsec:matching_CP}, we use the values $\alpha_{12}= 14.4^\circ$, $\rho = 0.3$ and $w = 5$, which are kept constant over the entire critical surface. 
We purposefully use a small $\alpha_{12}$ angle value because of the small physical quark masses, as motivated by Ref.~\cite{Pradeep:2019ccv}.

\subsection{TExS quantities}

\begin{figure}[!ht]
    \centering
    \includegraphics[width=\linewidth]{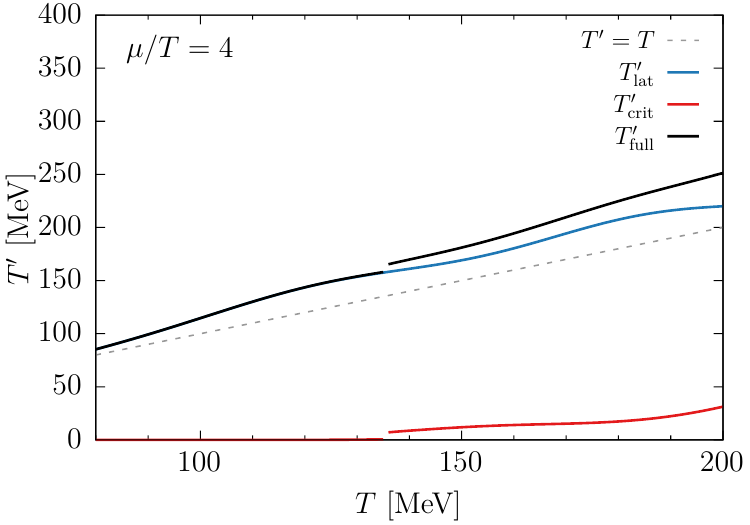}
    \caption{Breakdown of regular lattice $T^\prime$ expansion and critical contribution to $T^\prime_{full}(T)$ at $\mu/T = 4$, for average HIC conditions ($\langle n_S \rangle = 0$, $\langle n_Q/n_B \rangle = 0.4$).}
    \label{fig:Tprime_breakdown}
\end{figure}

We start by showing the effect of the critical contribution from the 3D Ising model when introduced into the TExS, first of all through the expanded temperature $T^\prime$. In Fig.~\ref{fig:Tprime_breakdown}, we plot $T^\prime_{\rm full}$ and its different components as functions of $T$, for average HIC conditions. 
$T^\prime_{crit}$ shows a discontinuity due to the first-order phase transition at around $T\approx136$ MeV, corresponding to $\mu_B \approx 520$ MeV, while being consistent with 0 below this temperature. This is expected, as we subtract $T^\prime_{\rm Taylor}$ when introducing the critical part through $T^\prime_{\rm Ising}$ in Eq.~\eqref{eq:Tfull}, to avoid double counting of low orders in the $T^\prime$ expansion (see also discussion in Sec.~\ref{subsec:matching_CP}).
This way, $T^\prime_{\rm full}$ only deviates from the original $T^\prime_\text{lat}$ after the transition line.

\begin{figure}[!ht]
    \centering
    \includegraphics[width=\linewidth]{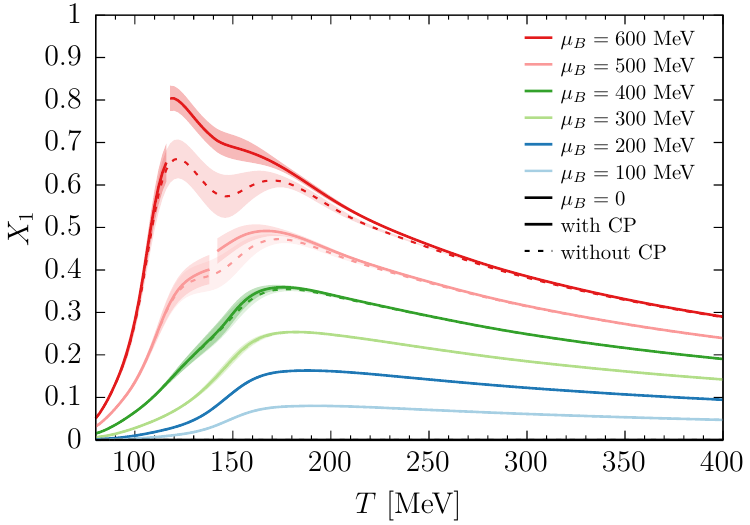}
    \caption{
    Comparison of dimensionless generalized charge density $X_1$ from the regular non-critical 4D-TExS (dashed lines) and from the new Ising 4D-TExS (full lines) at different values of finite $\mu_B$, for average HIC conditions ($\langle n_S \rangle = 0$, $\langle n_Q/n_B \rangle = 0.4$). 
    }
    \label{fig:X1_critical_vs_noncritical}
\end{figure}

In Fig.~\ref{fig:X1_critical_vs_noncritical}, we plot the dimensionless generalized charge density $X_1$ as a function of $T$ for fixed values of $\mu_B$ ranging from 0 to 600 MeV.
The results from regular 4D-TExS in average HIC conditions are shown by dashed lines with their corresponding error band, while the results with CP are showed by the continuous lines.
The results from the new 4D-TExS with CP stay very close to the regular non-critical 4D-TExS up until $\mu_B = 400$ MeV, and then show a discontinuity due to the first-order transition line in $X_1$. 
At higher $T$, both EoSs converge again, as the mapped Ising contribution fades away from the transition region.
Note that a few MeV-wide gap is observed at the transition due to the fact that we show results along the trajectory of average HIC conditions. Our numerical solver is not able to find a real solution that satisfies the imposed conditions when very close to the first-order hypersurface.
The oscillations which start to appear at high $\mu_B$ in the non-critical 4D-TExS seem to be stabilized when adding the critical contribution, although still present. They are symptomatic of the limitation of the $T^\prime$-expansion scheme, which is usually trusted up to $\mu/T = 3.5$ \cite{Abuali:2025tbd} but which we push to its limits here, as we reach values of $\mu / T \sim 5$ around the first-order transition at $\mu_B = 600$ MeV.

\begin{figure*}[!ht]
    \centering
    \includegraphics[width=\linewidth]{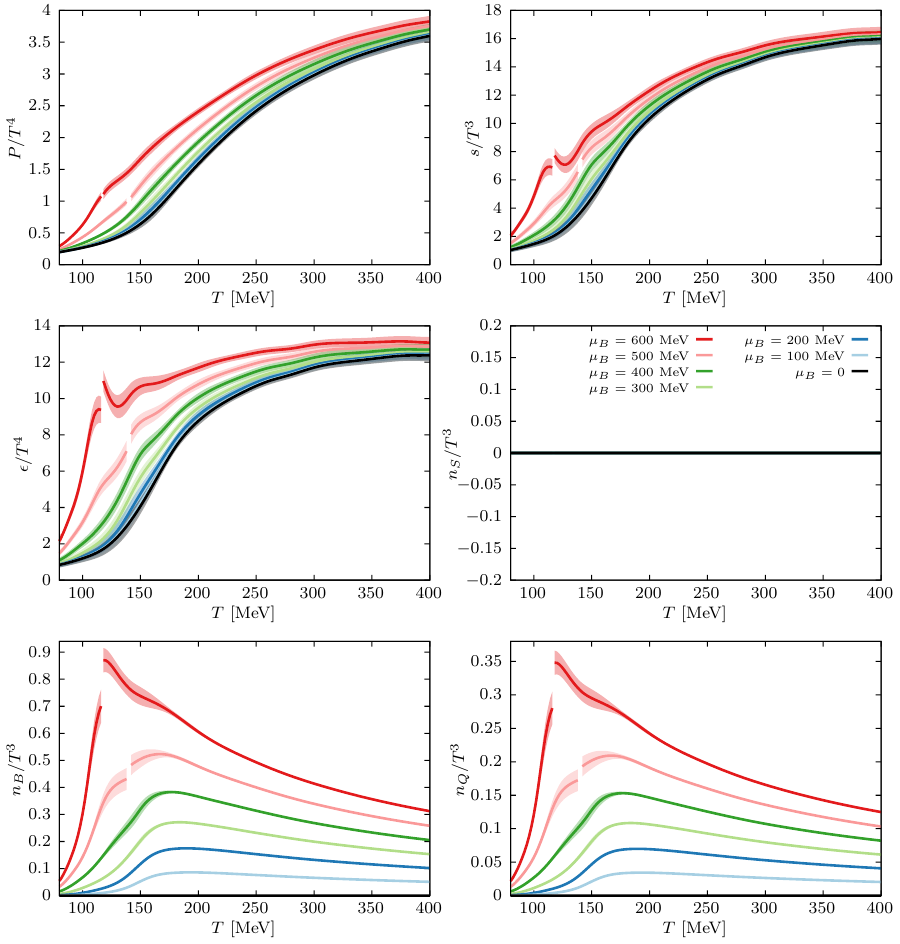}
    \caption{
    Dimensionless pressure and entropy, energy, strangeness, baryon and electric charge densities as functions of $T$ for different increasing values of $\mu_B$, for average HIC conditions ($\langle n_S \rangle = 0$, $\langle n_Q/n_B \rangle = 0.4$).
    } 
    \label{fig:thermo_SN}
\end{figure*}

\subsection{Thermodynamics}

We show results of normalized pressure, entropy density, energy density, net-strangeness density, net-baryon density and net-electric charge density as functions of $T$, and at multiple finite values of $\mu_B = 0$ -- 600 MeV in Fig.~\ref{fig:thermo_SN}.
Due to our choice to display average HIC condition results, the strangeness density $n_S$ is zero at all temperatures and values of $\mu_B$, while $n_Q = 0.4 \, n_B$. 

All quantities increase as expected with increasing $\mu_B$, and present a smooth evolution with $T$. 
We observe, however, the appearance of oscillations at $\mu_B = 600$ MeV in the entropy density as well. While not pathologic for the stability and causality of the EoS in the deconfined regime, we reserve for future works the task of merging the result of our expansion above the transition line, with the van der Waals Hadron Resonance Gas (vdW-HRG) EoS \cite{Vovchenko:2016rkn} in the hadronic phase, in order to have a proper description of QCD thermodynamics in the confined region. 
It has been shown that the TExS does not perform better than the Taylor expansion in describing the actual vdW-HRG EoS properly \cite{Kahangirwe:2024xyl}, which can also be expected based on the fact that the TExS relies on a different functional construction than vdW-HRG.

One can spot here again a gap due to the first-order transition starting at $\mu_B = 500$ MeV, which is also present in the pressure as we are not performing any mixed-phase construction in the transition region. In fact, we cannot access the spinodal branches in the 3D Ising model, as they lie in the complex plane \cite{An:2017brc}.

\begin{figure}[!h]
    \centering
    \includegraphics[trim={1mm 5mm 1mm 20mm}, width=\linewidth]{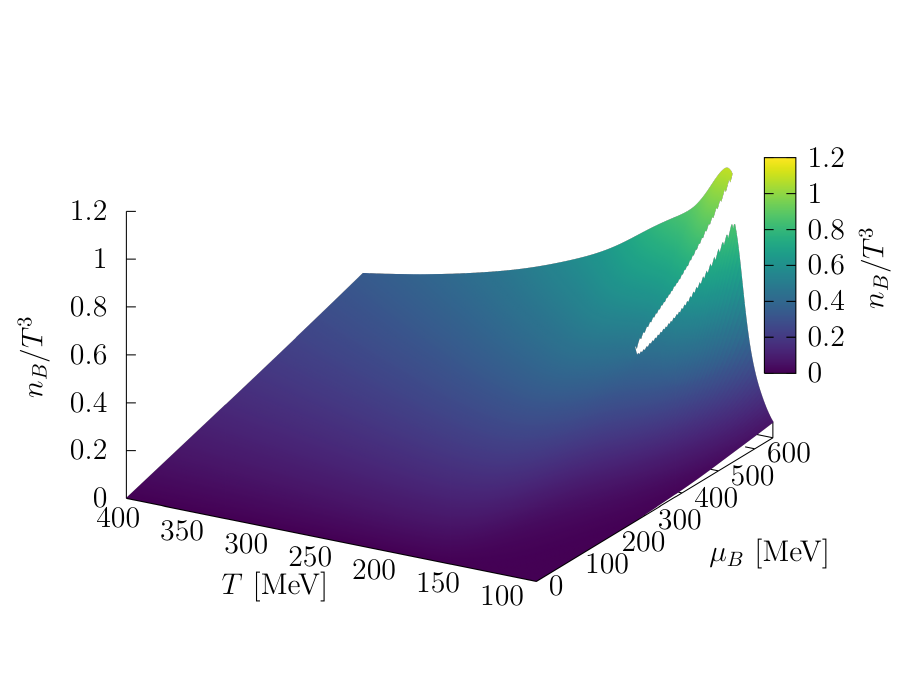}
    \centering
    \includegraphics[trim={1mm 5mm 1mm 20mm}, width=\linewidth]{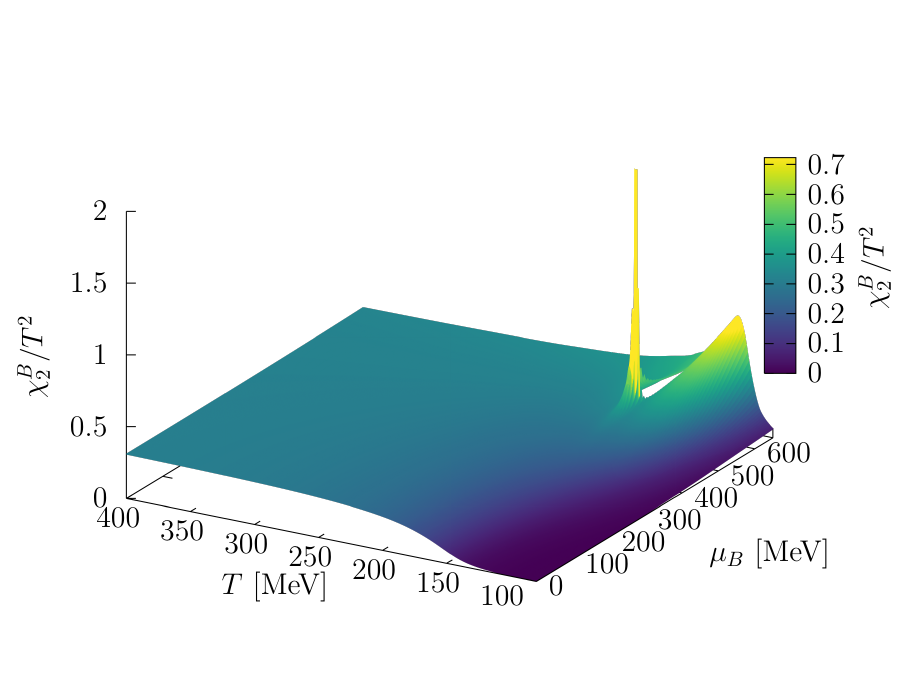}
    \caption{
    Three-dimensional profile of the normalized net-baryon density $n_B/T^3$ (\textit{top}) and normalized second-order baryon susceptibility $\chi_2^B/T^2$ (\textit{bottom}) as \mbox{functions} of $T$ and $\mu_B$, for average HIC conditions ($\langle n_S \rangle = 0$, $\langle n_Q/n_B \rangle = 0.4$).
    }
    \label{fig:nB+chi2B_vs_TmuB_3D}
\end{figure}

That same gap is made even more visible in the two panels of Fig.~\ref{fig:nB+chi2B_vs_TmuB_3D}, where the dimensionless baryon density $n_B$ and second-order baryon susceptibility $\chi_2^B$ are plotted in 3 dimensions as functions of $T$ and $\mu_B$. 
The second-order baryon susceptibility also shows a peak at the coordinates of the critical point, $(T_c,\mu_{B,c}) = (142, 492)$ MeV, illustrating the singularity caused by the latter.

The singularity of $\chi_2^B$ and, especially, the stronger singularities of the higher-order susceptibilities \cite{Stephanov:2008qz} are related to the fluctuation signatures of the critical point in beam-energy scan HIC experiments. More precisely, the measured (net-)proton multiplicity fluctuations can be related to the fluctuations of conserved quantities, such as baryon number, by maximum entropy method \cite{Pradeep:2022eil} -- in or out of equilibrium. While fluctuations of conserved quantities {\em in equilibrium} are given by susceptibilities, dynamical calculations are needed to take into account non-equilibrium effects on fluctuations, see e.g., recent work in Refs. \cite{Pradeep:2022mkf,Pihan:2026xng}.

\section{Conclusions}
\label{sec:conclusions}

In this article, we presented the first family of four-dimensional QCD equations of state describing the thermodynamics of nuclear matter at finite $(T,\mu_B,\mu_Q,\mu_S)$, based on a lattice QCD expansion and including a multi-dimensional critical contribution. 
We described how, from a procedure developed in the 2D space of finite $(T, \mu_B)$ \cite{Kahangirwe:2024cny}, we map the 3D-Ising model critical contribution to the QCD phase diagram, and insert it in the 4D-TExS formalism.
While the lattice 4D-TExS EoS we employ in our construction is left unchanged from Ref.~\cite{Abuali:2025tbd}, we update the way we embed the 3D Ising model critical contribution into it, making it fully consistent with the Stefan-Boltzmann limit normalization of the expansion.
Promoting the critical point to a critical surface in the full 4D thermodynamic space, as illustrated in Fig.~\ref{fig:critical_point_surface}, we defined a generic parametrization of the critical surface across the $(\mu_B, \mu_Q, \mu_S)$ subspace  based on a quadratic form. 
Once the generalized chemical potential coordinate of the critical point $\mu_c$ in a given direction is fixed, its temperature $T_c$ is automatically determined by construction, as we impose the critical point to sit on a line of constant $T^\prime$ equal to the pseudocritical temperature $T_0$ at vanishing chemical potentials. 

The use of a quadric form for the critical surface parameterization, presented in Sec.~\ref{subsec:critical_surface}, is motivated by charge-conjugation invariance of QCD, as well as exploratory works on the location of the critical point at finite $\mu_Q$ and $\mu_S$ from Refs.~\cite{Ueda:2013sia,Costa:2020dgc,Aryal:2021ojz}.
However, this formalism allows in principle to employ any possible parameterization, or even different functional forms or distributions for the critical surface, since the critical contribution from the 3D Ising model can be mapped independently for each different set of $(\theta, \varphi)$ angles onto the corresponding 2D $(T,\hmu)$ slice.
Along the same line, while we chose here to keep the mapping parameters $\alpha_{12}$, $\rho$ and $w$ constant across the critical surface, one can give them an explicit dependence on $(\theta, \varphi)$, or even employ a different mapping from 3D Ising to QCD as it is not universal.

We have shown our first results obtained with the Ising 4D-TExS construction, computed up to $\mu_B = 600$ MeV for average HIC conditions ($\langle n_S \rangle = 0$ and $\langle n_Q/n_B \rangle = 0.4$).
The critical surface was parametrized as a simple sphere for illustrative purpose, with a critical point located at $\mu_{B,c} = 492$ MeV, corresponding to a critical temperature of $T_c = 142$ MeV on this trajectory.
We observe a gap a few MeV-wide, marking the region where the conditions on the $S$ and $Q$ densities we impose do not have acceptable solutions. This is typical for the thermodynamically forbidden, or mixed-phase, region at a first-order phase transition.
While the obtained results show no pathological behavior, at least in the deconfined regime above the first-order transition, oscillations appear around the transition region at large $\mu_B$, due to the fact that we are pushing the $T^\prime$ expansion at large values of $\mu/T$ beyond the regime where it can be safely trusted.
We reserve for future work the merging of this EoS with the vdW-HRG one, in order to have a correct description of the thermodynamics of the hadronic phase, which is not properly captured by the TExS.

This new family of QCD equations of state represents a major advance toward studies of critical point effects using state-of-the-art heavy-ion collision simulations with three conserved charge evolution.
In the context of the BES-II results, its parametrizable nature makes it perfectly suitable for analysis pipelines comparing systematically theoretical inputs to experimental data \cite{Pihan:2026xng, Agarwal:2025ezo}. It allows one to vary the location and strength of the critical surface throughout the full $(T, \mu_B, \mu_Q, \mu_S)$ space, providing an appropriate tool to help infer the existence of a critical point in QCD matter.
\\

\begin{acknowledgments}
This material is based upon work supported by the National Science Foundation under grants No. PHY-2208724, PHY-2116686, PHY-2514763, PHY-2621752 and PHY-2623480, and within the framework of the MUSES collaboration, under Grant No. OAC-2103680. This material is also based upon work supported by the U.S. Department of Energy, Office of Science, Office of Nuclear Physics, under Award Numbers DE-SC0022023, DE-FG0201ER41195 and DE-SC0023861, as well as by the National Aeronautics and Space Administration (NASA) under Award Number 80NSSC24K0767.
\end{acknowledgments}
\appendix


\section{Transition-Line Slope}
\label{apdx:transition_slope}

In this Appendix, we derive the slope of the transition line defined by Eq.~\eqref{eq:CP_condition} at the critical point. This slope determines the angle $\alpha_1$ entering the Ising-to-4D-TExS mapping. In the $(T,\mu)$ plane, $\alpha_1$ is the angle between the transition line and the $\mu$ axis, such that
\begin{equation}
    \tan\alpha_1 = \left. \frac{dT_c}{d\mu} \right|_{Tc, \mu_c}.
\end{equation}

The derivation requires some care because the transformed temperature $T^\prime$ is naturally expressed as a function of $(T,\hmu)$, with
\begin{equation}
    \hmu = \frac{\mu}{T},
\end{equation}
whereas the slope is required in terms of the dimensional variables $(T,\mu)$. We therefore use \eqref{eq:CP_condition}, the condition on the transition line
\begin{equation}
    F(T,\mu) \equiv T^\prime-T_0 =
    T\left[1+\lambda_2(T)\frac{\mu^2}{T^2} \right]-T_0=0.
    \label{eq:F_def}
\end{equation}
Applying implicit differentiation along this curve gives
\begin{equation}
    \left. \frac{dT_c}{d\mu} \right|_{\rm line} =  - \frac{\left(\partial F/\partial\mu\right)_T}{
    \left(\partial F/\partial T\right)_\mu},
    \label{eq:slope_implicit}
\end{equation}
where the partial derivatives are evaluated on the transition line and, in particular, at $(T,\mu)=(T_c,\mu_c)$ when determining the critical-point slope.

We first evaluate the numerator. At fixed $T$,
\begin{equation}
    \left.\frac{\partial\hmu}{\partial\mu} \right|_T = \frac{1}{T},
\end{equation}
and therefore
\begin{equation}
    \left.\frac{\partial F}{\partial\mu}\right|_T 
    =\left. \frac{\partial T^\prime}{\partial\hmu}\right|_T \left. \frac{\partial\hmu}{\partial\mu}\right|_T
    =2\lambda_2(T)\hmu.
    \label{eq:dF_dmu}
\end{equation}

The denominator requires an additional contribution because $\hmu=\mu/T$ varies with $T$ when the dimensional chemical potential $\mu$ is held fixed. Using the chain rule,
\begin{equation}
    \left.\frac{\partial F}{\partial T}\right|_\mu = \left.\frac{\partial T^\prime}{\partial T}\right|_{\hmu}
    +
    \left.\frac{\partial T^\prime}{\partial\hmu} \right|_T \left. \frac{\partial\hmu}{\partial T} \right|_\mu.
    \label{eq:dF_dT_chain}
\end{equation}
The required derivatives are
\begin{equation}
    \left. \frac{\partial\hmu}{\partial T}\right|_\mu =  -\frac{\hmu}{T},
\end{equation}
\begin{equation}
    \left.\frac{\partial T^\prime}{\partial\hmu}  \right|_T  = 2T\lambda_2(T)\hmu,
\end{equation}
and
\begin{equation}
    \left. \frac{\partial T^\prime}{\partial T} \right|_{\hmu}  = 1 +  \hmu^2 \left[\lambda_2(T)
    + T\lambda_2^\prime(T)\right].
\end{equation}
Substituting these expressions into Eq.~\eqref{eq:dF_dT_chain} yields
\begin{align}
    \left.\frac{\partial F}{\partial T}\right|_\mu
    &= \underbrace{\left[1 + \hmu^2\bigl(\lambda_2(T) + T\,\lambda_2^\prime(T)\bigr)\right]}_{(\partial T^\prime/\partial T)_{\hmu}}
       \;+\; (2T\,\lambda_2\,\hmu)\,\Bigl(-\frac{\hmu}{T}\Bigr) \nonumber\\
    &= 1 + \hmu^2\bigl[T\,\lambda_2^\prime(T) - \lambda_2(T)\bigr] \, .
    \label{eq:dF_dT}
\end{align}
The change in the coefficient of $\lambda_2(T)$ relative to
$\left(\partial T^\prime/\partial T\right)_{\hmu}$
arises entirely from the $T$ dependence of $\hmu=\mu/T$ at fixed $\mu$.

Combining Eqs.~\eqref{eq:dF_dmu} and \eqref{eq:dF_dT} with Eq.~\eqref{eq:slope_implicit}, and evaluating the result at the critical point, gives
\begin{equation}
    \left. \frac{dT_c}{d\mu}\right|_{T_c, \mu_c} = - \frac{ 2\lambda_2(T_c)\hmu_c }{ 1 + \hmu_c^2 \left[T_c\lambda_2^\prime(T_c) - \lambda_2(T_c) \right]}.
    \label{eq:dTc_dmu}
\end{equation}
The corresponding mapping angle is therefore
\begin{equation}
    \tan\alpha_1 = \left| -\frac{ 2\lambda_2(T_c)\hmu_c }{ 1 + \hmu_c^2 \left[ T_c\lambda_2^\prime(T_c) -\lambda_2(T_c)\right] }
    \right|.
    \label{eq:alpha1}
\end{equation}
For the parameter range considered here, this expression reduces to the form used in Eq.~\eqref{eq:slope}.

\section{Ising density}
\label{apdx:Ising_density}

Throughout this appendix, the direction $(\theta,\varphi)$ is fixed and its indices are suppressed, as in Sec.~\ref{subsec:ising-4dtexs-mapping}. We write $X_{2,{\rm lat}}$ for the zero-density lattice susceptibility denoted by $X_2$ in Eq.~\eqref{eq:4dtexs_definition}, the subscript distinguishing it from $X_{2,{\rm Ising}}$. The density entering this relation is the derivative of the normalized Ising pressure introduced in Sec.~\ref{subsec:Ising_scaling},
\begin{equation}
    n_{\rm Ising}(T,\hmu)
    \equiv
    \left.\frac{\partial(P_{\rm Ising}/T^4)}{\partial\hmu}
    \right|_{T,\theta,\varphi}.
    \label{eq:nIsing_definition}
\end{equation}
Thus $n_{\rm Ising}$ is dimensionless and refers to the generalized chemical potential. Along the $\mu_B$ axis, it corresponds to the mapped Ising baryon density divided by $T^3$. At fixed $T$, $\partial_{\hmu}=T\partial_\mu$, so the chain rule gives
\begin{equation}
\begin{aligned}
    n_{\rm Ising}=T\biggl[
    &\left.\frac{\partial(P_{\rm Ising}/T^4)}{\partial r}\right|_{T,h}
      \left.\frac{\partial r}{\partial\mu}\right|_T
    \\&\quad 
    +\left.\frac{\partial(P_{\rm Ising}/T^4)}{\partial h}\right|_{T,r}
      \left.\frac{\partial h}{\partial\mu}\right|_T
    \biggr].
    \label{eq:nIsing_chain}
\end{aligned}
\end{equation}
The pressure derivatives are evaluated using the parametric representation $(R,\Theta)$. Differentiating the mapping in Eq.~\eqref{eq:mapping}, with its critical-point parameters held fixed, yields:
\begin{align}
    \left.\frac{\partial h}{\partial\mu}\right|_T
    &=-\frac{2\lambda_2(T)\hmu}
    {w' T_c\,\partial_T T'_c\,\sin\alpha'_{12}},
    \label{eq:appB_h_derivative}\\
    \left.\frac{\partial r}{\partial\mu}\right|_T
    &=-\frac{1}{\rho'}\left[
    \frac{\mu}{\mu_c \, w'\,T_c}
    +\cos\alpha'_{12}\left.\frac{\partial h}{\partial\mu}\right|_T
    \right].
    \label{eq:appB_r_derivative}
\end{align}
Here we have used the truncation $T'_{\rm lat}=T[1+\lambda_2(T)\hmu^2]$, and the parameters $w'$, $\rho'$ and $\alpha'_{12}$ are those of Eq. \eqref{eq:mapping}, obtained from $w$, $\rho$ and $\alpha_{12}$ through Eq. \eqref{eq:linear-relations}.

Equations~\eqref{eq:nIsing_chain}--\eqref{eq:appB_r_derivative} specify the density obtained from the mapped Ising pressure. To relate this density to a temperature shift, we expand the zero-density lattice susceptibility around $T'=T_0$:
\begin{equation}
\begin{aligned}
    X_{2,{\rm lat}}(T',0)
    ={}&X_{2,{\rm lat}}(T_0,0)\\
    &+\left.\frac{\partial X_{2,{\rm lat}}}{\partial T}\right|_{T_0}
    (T'-T_0)
    +\mathcal{O}\!\left((T'-T_0)^2\right).
    \label{eq:X2_linearization}
\end{aligned}
\end{equation}
The expansion point is $T_0$ because the matching has to hold near the transition line, where the Ising sector carries the critical structure. Indeed, according to Eq.~\eqref{eq:mapping}, $h\propto T'_{\rm lat}-T_0$ vanishes on that line, and $T'_{\rm lat}(T_c,\hmu_c)=T_0$ at the critical point.
At fixed $\hmu$, the SB prefactor $\overline X_1(\hmu)/\overline X_2(0)$ in Eq.~\eqref{eq:4dtexs_definition} is unchanged by a shift of $T'$. Applying Eq.~\eqref{eq:X2_linearization} to $T'_{\rm lat}+T'_{\rm Ising}$ and $T'_{\rm lat}$, and subtracting, therefore gives, to linear order,
\begin{equation}
\begin{aligned}
    \frac{\overline X_1(\hmu)}{\overline X_2(0)}
    \bigl[X_{2,{\rm lat}}(T'_{\rm lat}
            &+T'_{\rm Ising},0)
            -X_{2,{\rm lat}}(T'_{\rm lat},0)\bigr]\\
    &\quad\simeq
    \frac{\overline X_1(\hmu)}{\overline X_2(0)}
    \left.\frac{\partial X_{2,{\rm lat}}}{\partial T}\right|_{T_0}
    T'_{\rm Ising}.
    \label{eq:appB_density_shift}
\end{aligned}
\end{equation}
Before the low-order Taylor subtraction, we match this leading density response to $n_{\rm Ising}$. Finally, solving for the shift motivates the prescription:
\begin{equation}
    T'_{\rm Ising}(T,\hmu)
    \equiv
    \left(\left.\frac{\partial X_{2,{\rm lat}}}{\partial T}\right|_{T_0}\right)^{-1}
    \frac{n_{\rm Ising}(T,\hmu)}
    {\overline X_1(\hmu)/\overline X_2(0)},
    \label{eq:appB_TIsing_matching}
\end{equation}
which is nothing but Eq.~\eqref{eq:Tcrit-critical-point}. 
$T'_{\rm Ising}$ is a shift of $T'$, added to $T'_{\rm lat}$ in Eq. \eqref{eq:Tfull}, which is why no $T_0$ appears in Eq. \eqref{eq:appB_TIsing_matching}. The ratio at $\hmu=0$ is understood by its continuous limit, using Eqs.~\eqref{eq:nIsing_taylor} and \eqref{eq:SB_prefactor}.

The linearization motivates the matching prescription, which is then used in the full nonlinear TExS relation. Therefore, $n_{\rm Ising}$ denotes the mapped Ising density used to construct the shift.

\bibliographystyle{apsrev4-2}
\bibliography{references}

\end{document}